\documentclass[micromachines,article,submit,moreauthors,pdftex,10pt,a4paper]{mdpi_noline}
\firstpage{1}
\articlenumber{x}
\doinum{10.3390/------}
\pubvolume{xx}
\pubyear{2016}
\copyrightyear{2016}
\externaleditor{Academic Editor: Dr. Marc Desmulliez, Anne Bernassau and Baixin Chen}
\history{Received: date; Accepted: date; Published: date}
\usepackage{bm}
\usepackage{siunitx}
\usepackage{subcaption}
\newcommand{\dm}{\mathrm{d}}
\renewcommand{\vec}[1]{\bm{#1}}
\newcommand{\ten}[1]{{\bm{#1}}}

\newcommand{\grad}{\vec{\nabla}}
\newcommand{\divv}{\vec{\nabla} \!\cdot\!}
\newcommand{\partt}{\partial_t}
\newcommand{\us}{\vec{u}_{\mathrm{s}}}

\newcommand{\sig}{\ten{\sigma}}
\renewcommand{\tau}{\ten{\tau}}
\newcommand{\sigf}{\sig_{\mathrm{f}}}

\newcommand{\sigs}{\sig_{\mathrm{s}}}
\renewcommand{\eqref}[1]{Eq.~(\ref{eq:#1})}

\newcommand{\figref}[1]{Fig.~\ref{fig:#1}}

\newcommand{\iot}{\ii\omega t}
\newcommand{\ii}{\mathrm{i}}
\newcommand{\ee}{\mathrm{e}}

 \theoremstyle{mdpi}
 \newcounter{thm}
 \newcounter{ex}
 \newcounter{re}
 \theoremstyle{mdpidefinition}

\Title{Modeling of microdevices for SAW-based acoustophoresis --- a study of boundary conditions}

\Author{Nils Refstrup Skov  and Henrik Bruus*}
\AuthorNames{Nils Refstrup Skov and Henrik Bruus}

\address[1]{%
Department of Physics, Technical University of Denmark,\\
DTU Physics Building 309, DK-2800 Kongens Lyngby, Denmark.}

\corres{Correspondence: nilsre@fysik.dtu.dk and bruus@fysik.dtu.dk; Tel.: +45 4525-3307}

\abstract{
We present a finite-element method modeling of acoustophoretic devices consisting of a single, long, straight, water-filled microchannel surrounded by an elastic wall of either borosilicate glass (pyrex) or the elastomer polydimethylsiloxane (PDMS) and placed on top of a piezoelectric transducer that actuates the device by surface acoustic waves (SAW). We compare the resulting acoustic fields in these full solid-fluid models with those obtained in reduced fluid models comprising of only a water domain with simplified, approximate boundary conditions representing the surrounding solids. The reduced models are found to only approximate the acoustically hard pyrex systems to a limited degree for large wall thicknesses and not at all for the acoustically soft PDMS systems.}


\begin{document}

%%%%%%%%%%%%%%%%%%%%%%%%%%%%%%%%%%%%%%%%%%
%% Sections that are not mandatory are listed as such. The section titles given are for Articles. Review papers and other article types have a more flexible structure.

%% Only for the journal Gels: Please place the Experimental Section after the Conclusions

%%%%%%%%%%%%%%%%%%%%%%%%%%%%%%%%%%%%%%%%%%

\section{Introduction} \label{sect:Intro}
Separation of particles and cells is important in a wide array of biotechnological applications \cite{Petersson2007, Amini2014, Shi2009, Shi2011, Chen2014, Lee2015, Liga2015}. This has traditionally been carried out by bulk processes including centrifugation, chromatography, and filtration. However, during the last three decades, microfluidic devices have proven to be a valuable alternative \cite{Pamme2007, Petersson2007, Liga2015}, as they allow for lower sample sizes and decentralized preparations of biological samples, increasing the potential for point-of-care testing. Microfluidic methods for separating particles suspended in a medium include passive methods where particle separation is solely determined by the flow and the size or density of particles \cite{Sethu2006, Huh2007, Amini2014, Sugiyama2014, Zhang2016}, and active methods where particles migrate due to the application of various external fields  each targeting specific properties for particle sorting \cite{Petersson2007, Pamme2006, Shi2009, Shi2011, Guldiken2012, Travagliati2013, Lee2015, Guo2016}. Acoustophoresis is an active method, where emphasis is on gentle, label-free, precise handling of cells based on their density and compressibility relative to the suspension medium as well as their size \cite{Bruus2011c}. Within biotechnology, acoustophoresis has been used to confine, separate, sort or probe particles such as microvesicles~\cite{Evander2015, Lee2015}, cells~\cite{Petersson2007, Wiklund2012b, Collins2015, Ahmed2016, Guo2016, Augustsson2016}, bacteria~\cite{Hammarstrom2012, Carugo2014}, and biomolecules \cite{Sitters2015}. Biomedical applications include early detection of circulating tumor cells in blood~\cite{Augustsson2012, Li2015} and diagnosis of bloodstream infections~\cite{Hammarstrom2014a}.

\begin{figure}[t]
\centering
\includegraphics[width=\columnwidth]{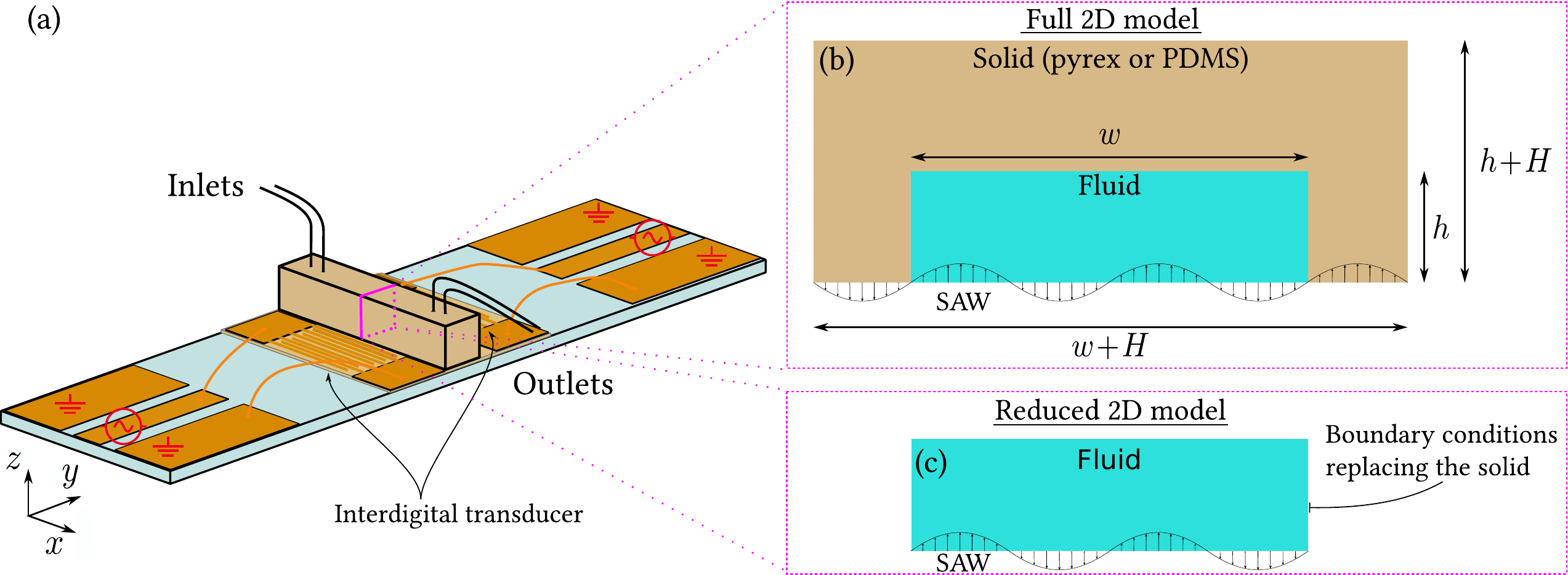}%
\caption{\label{fig:Chip} (a) Sketch of the generic acoustophoretic device under study. A fluid flows through a long straight microchannel defined by a surrounding solid wall (pyrex or PDMS, light brown) and placed on top of a piezoelectric substrate (light blue). By actuating of interdigital transducers (IDTs, dark brown) placed on either side of the device, surface acoustic waves (SAW) propagate along the surface of the substrate, and when timed properly they form a standing wave. (b) Sketch in the transverse $yz$ cross section of the full 2D model consisting of a solid domain with wall thickness $H$ and a fluid domain of width $w$ and height $h$ . (c) Similar sketch of the reduced 2D model, which consists solely of the fluid domain in (b), but with boundary conditions (hard wall or lossy wall) representing in an approximate manner the surrounding solid.}
\end{figure}

The acoustic fields used in acoustophoresis are mainly one of the following two kinds: (1) Bulk acoustic waves (BAW), which are set up in the entire device and used in systems with acoustically hard walls. BAW depend critically on the high acoustic impedance ratio between the walls and the water. (2) Surface acoustic waves (SAW), which are defined by interdigital electrodes on the piezoelectric transducer and propagate along the transducer surface. SAW are nearly independent of the acoustical impedance ratio of the device walls and the microchannel, and this feature makes the SAW technique versatile. SAW can be used both with hard- and soft-walled acoustophoretic devices, often in the generic setup sketched in \figref{Chip}, where the fluid-filled microchannel is encased by a solid material and is placed directly on top of the piezoelectric substrate to ensure optimal coupling to the SAW induced in the substrate.

Because SAW-based acoustophoretic microdevices are very promising as powerful and versatile tools for manipulation of microparticles and cells, numerical modeling of them are important, both for improved understanding of the acoustofluidic conditions within the devices and to guide proper device design. In the literature, such modeling has been performed in numerous ways. For many common elastic materials, the dynamics of the walls are straightforward to compute fully through the usual Cauchy model of their displacement fields $\vec{u}$ and stress tensors $\vec{\sigma}$. The coupling to the acoustic pressure $p$ and velocity $\vec{v}$ in the microchannel, described by the Navier--Stokes equation, is handled by the continuity conditions $\vec{n}\cdot\vec{\sigma}_{\mathrm{s}} = \vec{n}\cdot\vec{\sigma}_{\mathrm{f}}$ and $\partial_t\vec{u} = \vec{v}$ of the stress and velocity fields. This full model is discussed in detail in Section \ref{sect:Numeric}. For acoustically hard walls, such as borosilicate glass (pyrex) with a high impedance ratio ($\tilde{Z} = 8.4$) relative to water, the full model is often replaced by a reduced model (exact for $\tilde{Z} = \infty$) with less demanding numerics, where only the fluid domain in the microchannel is treated, and where the elastic walls are replaced by the so-called hard-wall boundary condition demanding zero acoustic velocity at the boundary of the fluid domain \cite{Muller2012, Leibacher2014, Muller2015}. For rubber-like polymers such as the often used PDMS, the full device modeling is more challenging. For large strains (above 40~\%), a representation of the underlying macromolecular network of polymer chains is necessary \cite{Arruda1993}, while for the moderate strains appearing in typical acoustophoretic devices, standard linear elasticity suffices \cite{Yu2009,Bourbaba2013}. Some authors argue that the low ratio of the transverse to longitudinal speed of sound justifies a fluid-like model of PDMS based on a scalar Helmholtz equation \cite{Leibacher2014, Darinskii2016}. Furthermore, since the acoustic impedance ratio $\tilde{Z} = 0.7$ between PDMS and water is nearly unity, the full model has in the literature been replaced by a reduced model, consisting of only the fluid domain with the so-called lossy-wall boundary condition condition representing in an approximate manner the acoustically soft PDMS walls \cite{Nama2015, Mao2016, Guo2016}.

The main aim of this paper is to investigate to which extent the numerically less demanding hard- and lossy-wall reduced models compare with the full models for SAW-based acoustofluidic devices. In the full models we study the two generic cases of acoustically hard pyrex walls and acoustically soft PDMS walls, both treated as linear elastic materials. In the reduced models, the pyrex and PDMS walls are represented by hard-wall and lossy-wall boundary conditions, respectively. In all the models, the fluid (water) is treated as a Newtonian fluid governed by the continuity equation and the Navier--Stokes equation. Our main result is, that for pyrex walls the reduced model approximates the full model reasonably well for sufficiently thick walls, but fails for thin walls, while for PDMS walls the lossy-wall boundary condition fails regardless of the wall thickness.

%%%%%%%%%%%%%%%%%%%%%%%%%%%%%%%%%%%%%%%%%%

\begin{figure}[t]
\centering
\includegraphics[]{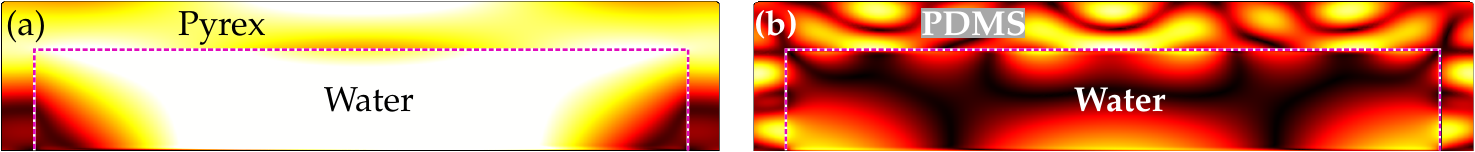}
\caption{\label{fig:FullModel} Examples of full model results for $h =\SI{125}{\micro\meter}$, $w =\SI{600}{\micro\meter}$, and $ H = \SI{60}{\micro\meter}$, see \figref{Chip}(b). (a)~Color plot from 0 mm/s (black) to 1.5 mm/s (white) of the velocity field $\big|-\ii \omega \us\big|$ in the pyrex and $\big|\vec{v}_{\mathrm{f}}\big|$ in the water obtained in a full-model simulation of a pyrex SAW device actuated at the on-resonance frequency $f_\mathrm{res} = 1.24$~MHz. (b) The same as in (a) but for a full-model PDMS SAW device. The white-magenta dashed line indicates the solid-fluid interface.}
\end{figure}

\section{Results: comparing the full and reduced 2D models} \label{sect:Results}

In the following, we present our results for the numerical simulations of the acoustic fields in the reduced and full models with SAW actuation, and we compare the two cases. As the microchannels are long and straight along the $x$ direction, we assume translational invariance along $x$ and restrict the calculational domain to the two-dimensional (2D) cross section in the $yz$ plane. The full model consists of coupled fluid and solid domains, whereas the reduced model consists of a single fluid domain with boundary conditions that in an approximate manner represent the walls. The principle of our model approach is illustrated in \figref{Chip}, while the models are described in detail in Section \ref{sect:Numeric}.

\subsection{Pyrex devices: full model and reduced hard-wall model} \label{sect:HardWall}
We consider first the full model of a pyrex microdevice, in which a rectangular water-filled channel of width $w$ and height $h$ is encased by a pyrex wall of height $h+H$ and width $w+H$, see \figref{Chip}(b). We simulate the case of actuating the system both at the horizontal standing half-wave resonance in the water $f_\mathrm{res} = c_0/2w = \SI{1.24}{\mega\hertz}$ often exploited in experiments, and at the off-resonance frequency $f_\mathrm{off}  = \SI{6.65}{\mega\hertz}$ chosen to facilitate comparisons with the literature \cite{Nama2015}. An example of a full-model result for the velocity field $- \ii \omega \vec{u}$ in the pyrex and $\vec{v}_{\mathrm{f}}$ in the water, is shown in \figref{FullModel}(a).

\begin{figure}[t]
\centering
\includegraphics[width=\columnwidth]{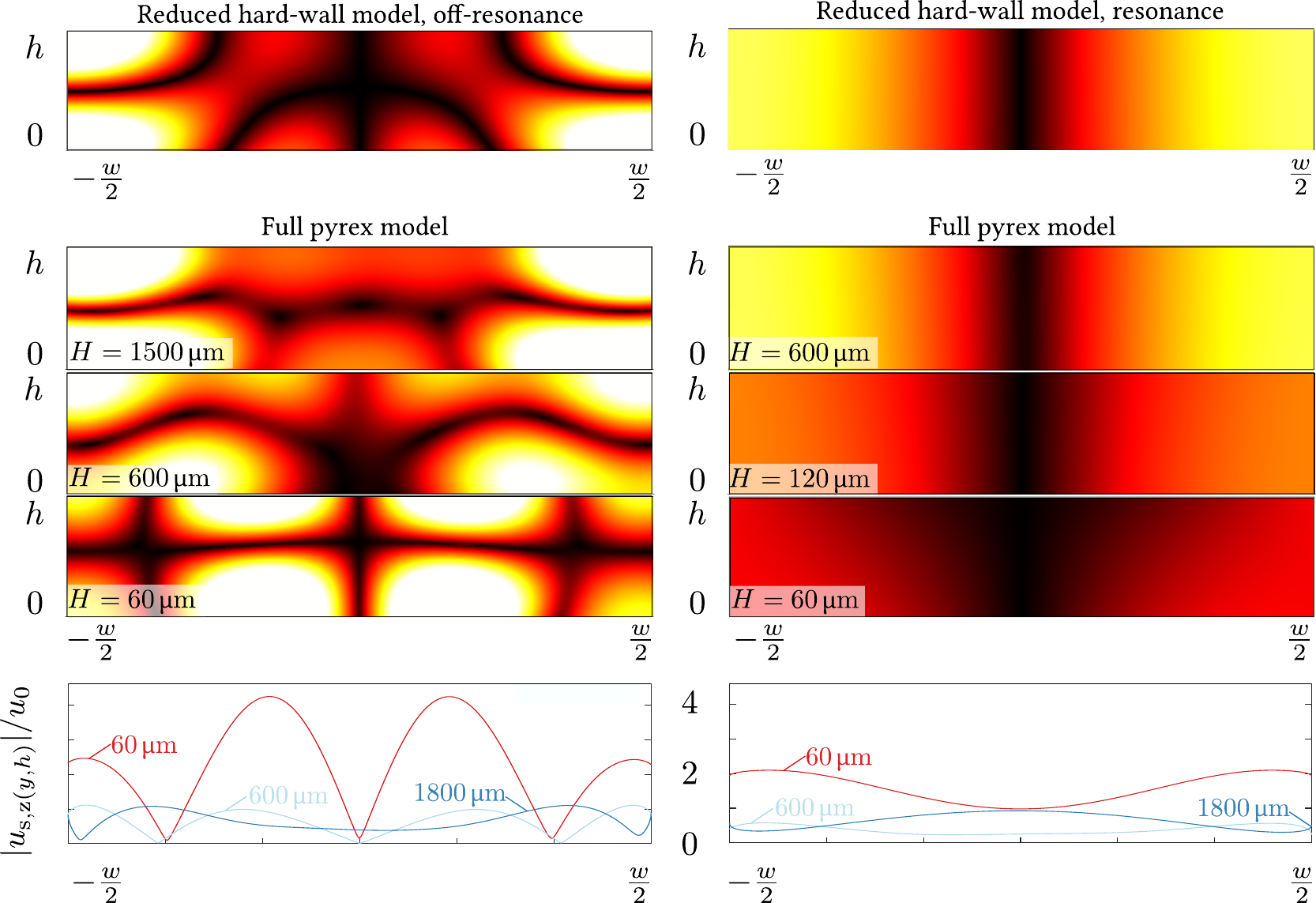}
\caption{\label{fig:AllHard} Left column: color plots from 0~kPa (black) to 40~kPa (white) at the off-resonance frequency $f_{\mathrm{off}} = \SI{6.65}{\mega\hertz}$ of the amplitude $|p_\mathrm{f}|$ of the first-order pressure field $p_\mathrm{f}$ in the fluid domain of the reduced hard-wall model and the full pyrex model \figref{Chip}(a), but with $H = 60$, 600, and \SI{1500}{\micro\metre}. The surrounding pyrex is not shown. Right column: the same as to the left, but at the on-resonance actuation frequency $f = \SI{1.24}{\mega\hertz}$ and with the color plots ranging from 0~kPa (black) to 80~kPa (white). Bottom row: off- and on-resonance line plots of the amplitude $|u_{s,z}(y,h)|$ of the vertical displacement along the top fluid-solid interface normalized by the amplitude $u_0$ of the SAW actuation displacement for wall thickness $H = 60$, 600, and \SI{1500}{\micro\metre}.}
\end{figure}

We then investigate to which extent the full model can be approximated by the reduced hard-wall model often used in the literature \cite{Bruus2012,Muller2012}, where only the water domain is considered, while the pyrex walls are represented by the hard-wall condition. In \figref{AllHard} we show for both off-resonance (left column)  and on-resonance (right column) actuation, a qualitative comparison between the reduced and the full model, with wall thickness $H$ ranging from \SIrange{60}{1800}{\micro\metre}. Considering the resulting amplitudes $|p_\mathrm{f}|$ of first-order pressure field $p_\mathrm{f}$ in the water domain, we note that for off-resonance actuation at the frequency $f_\mathrm{off}$, the full model with thick walls $H = \SI{1500}{\micro\metre}$ has some features in common with the reduced model. There are pressure anti-nodes in the corners and an almost horizontal pressure node close to the horizontal centerline. For decreasing wall thickness $H$ in the full model, the pressure field changes qualitatively, as the pressure anti-nodes detach from the side walls and shift towards the center of the fluid domain. When actuated on resonance at the frequency $f_\mathrm{res}$,  for wall thicknesses as low as $H = \SI{120}{\micro\metre}$, the full-model pressure is nearly indistinguishable from that of the hard-wall reduced model, namely a cosine function with vertical pressure anti-nodal lines along the side walls and a vertical pressure nodal line in the center. For the smallest wall thickness $H = \SI{60}{\micro\metre}$ the iso-bars in the full model tilts relative to vertical. In summary, the correspondence between the full and the reduced model is overall better for on-resonance actuation, but for a large wall thickness the reduced hard-wall model describes the full pyrex model reasonably well.

Finally, in the bottom row of \figref{AllHard}, we investigate for the full pyrex model model the displacement at the upper boundary in units of the imposed displacement amplitude $u_0$ at the SAW-actuated lower boundary. If the hard-wall condition of the reduced model is good, this displacement should be very small. However, from the figures it is clearly seen that for the thin wall $H = \SI{60}{\micro\metre}$, the upper-wall displacement is significant, with an amplitude of $4u_0$ at $f_\mathrm{off}$ and $2u_0$ at $f_\mathrm{res}$. As the wall thickness $H$ increases, the upper-wall displacement amplitudes decreases towards $u_0$. Again, this reflects that the reduced hard-wall model is in fair agreement with the full model for a large wall thickness $H$, and it is better on resonance, where the specific values at the boundaries are less important as the pressure field is dominated by the pressure eigenmode that does in fact fulfill the hard-wall condition, see Sec. \ref{sect:Resonance}.

\begin{figure}[t]
\centering
\includegraphics[width=\columnwidth]{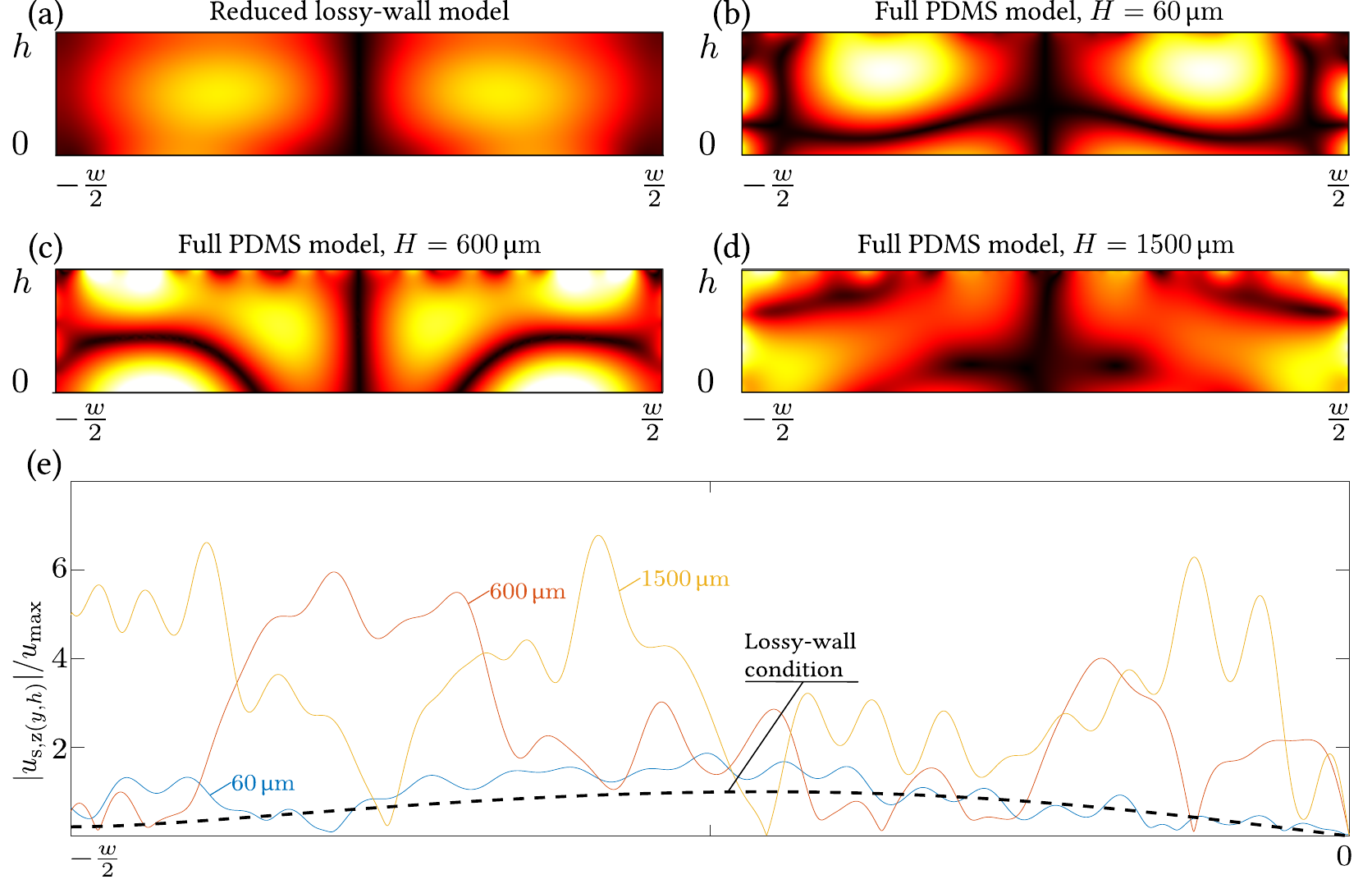}
\caption{\label{fig:4Lossy}
Color plot from 0 kPa (black) to 30 kPa (white) of the amplitude $|p_\mathrm{f}|$ of the first-order pressure field $p_\mathrm{f}$ in the fluid domain of (a) the reduced lossy-wall model and (b)-(d) the full PDMS model with wall thickness $H = 60$, 180, and \SI{1500}{\micro\metre}. The surrounding PDMS is not shown.
(e)~Line plots of the normalized amplitude $|u_{s,z}(y,h)|/u_{\mathrm{max}}$ of the vertical displacement along the upper fluid-solid interface in the full model (full colored lines) with $H = 60$, 600, and \SI{1500}{\micro\metre} and in the reduced lossy-wall model (dashed black line). The normalization unit $u_{\mathrm{max}}$ is the maximum amplitude found in the reduced lossy-wall model.}
\end{figure}

\subsection{PDMS devices: full model and reduced lossy-wall model} \label{sect:LossyWall}
We then move on to show the same comparisons, but where the full model has PDMS walls, and the reduced model has the lossy-wall boundary condition, which takes deformation in the normal direction of the wall into account. The reduced lossy-wall model for PDMS, actuated at the off-resonance frequency $f_\mathrm{off} = 6.65$~MHz, is exactly the one used by Nama \textit{et al.}~\cite{Nama2015}. Given the low impedance ratio $\tilde{Z} = 0.7$ between PDMS and water, there is no resonance, so we do not show any results for actuation at $f_\mathrm{res}$. The results for reduced lossy-wall model and the full PDMS model is shown in \figref{4Lossy} with plots similar to the ones in the left column of \figref{AllHard} for the reduced hard-wall model and the full pyrex model.

Initially, we compare in \figref{4Lossy}(a)-(d) the amplitude $|p_\mathrm{f}|$ of the first-order pressure field $p_\mathrm{f}$ of the reduced lossy-wall model with that of the full PDMS model for the wall thickness $H$ varying from \SIrange{60}{1500}{\micro\metre}.
Due to the lossy-wall boundary condition~(\ref{eq:LossyWall}), the ellipsoidal pressure anti-nodes in \figref{4Lossy}(a) traverse the fluid domain upwards during one oscillation cycle. This is in stark contrast to the pressure structures of the full PDMS model in \figref{4Lossy}(b)-(d), which are stationary due to the free stress condition~(\ref{eq:BCsFREE}) imposed on the exterior of the PDMS.
Moreover, the pressure structure of the reduced lossy-wall model consists of only two pressure antinodes, which is much simpler than the multi-node structure of the full PDMS model. In fact, the only common feature in the pressure fields is the appearance of a well-defined pressure node along the vertical centerline.

The poor qualitative agreement between the pressure field in the reduced lossy-wall model and in the full PDMS model is further supported in \figref{4Lossy}(e), where the upper-wall displacement amplitudes of the models are shown. We introduce the unit $u_{\mathrm{max}}$ as the maximum displacement along the upper-wall in the reduced lossy-wall model, and not that the lossy-wall condition imposes a broad single-node sinusoidal velocity amplitude of unity magnitude, while each of the four full model cases ($H = 60$, 600, and \SI{1500}{\micro\metre}) shows a more erratic multi-peaked velocity amplitude of magnitudes ranging from 2 to 6.

%%%%%%%%%%%%%%%%%%%%%%%%%%%%%%%%%%%%%%%%%
\section{Discussion} \label{sect:Discussion}

\subsection{Physical limitations of the hard-wall condition} \label{sect:HardDiscussion}
As illustrated in \figref{AllHard} there are clear discrepancies between the fields obtained by the reduced hard-wall model and those found using the full pyrex models. This can likely be attributed to two factors in particular: the finite stiffness and density of pyrex, and the non-local SAW actuation imposed along the bottom edge in the model.

The hard-wall condition is physically correct for an infinitely stiff and dense wall, which does not undergo any deformation or motion regardless of the stress exerted by the fluid. A hard wall thus reflects all acoustic energy incident on it back into the fluid. However, pyrex has a finite stiffness and density, it will thus deform and allow for a partial transmittance of acoustic energy from the fluid. This aspect is part of the full pyrex model, but not of the reduced hard-wall model.

The specific SAW actuation is also different in the full and the reduced model. The microdevice rests on top of the piezoelectric substrate, so in the full model, the standing SAW along the surface of the piezoelectric substrate (typically lithium niobate) will transmit significant amounts of acoustic energy directly into both the pyrex wall and the water, but only the latter is taken into account in the reduced hard-wall model. The coupling between lithium niobate and pyrex is strong since the direction-dependent elastic stiffness coefficients of lithium niobate lies in the range from 53 to \SI{200}{\giga\pascal} \cite{Weis1985} and the Young's modulus of pyrex of \SI{64}{\giga\pascal} lies in the same range \cite{Narottam1986}. Consequently, the interface between the pyrex wall and the water will move under the combined action of the acoustic fields loaded into the pyrex and the water, respectively.

\begin{table}[t]%,
\centering
\caption{\label{tab:Parameters} List of parameters used for geometry, materials, and SAW in the numerical model.}
\begin{tabular}{lcccccc}
\textbf{Quantity} & \textbf{Symbol}	& \textbf{Unit} 	& \textbf{Pyrex}		& \textbf{PDMS} & \textbf{Water}  & \textbf{SAW}    \\
 & & & \cite{Narottam1986} & \cite{Madsen1983,Zell2007} & \cite{Muller2014} & \cite{Nama2015} \\
														\midrule
Width & $\frac12 H$ or $w$ & \SI{}{\micro\metre}  & 30 -- 900	& 30 -- 750 & 600 & -\\	
Height & $H$ or $h$  & \SI{}{\micro\metre}  & 60 -- 1800	& 60 -- 1500 & 125 & -\\	
														\midrule	
Density &	$\rho_{\mathrm{f}}$ or $\rho_\mathrm{s}$ & \SI{}{\kilo\gram\per\cubic\metre} 	& 2230 & 1070 & 997 & -\\
Longitudinal sound speed &	$c_{\rm{L}}$ or $c_0$ & \SI{}{\metre\per\second}		& 5591	& 1030 & 1496& -\\
Transversal sound speed & $c_{\rm{T}}$ & \SI{}{\metre\per\second}	&	3424& 100 & - & -\\
Acoustic impedance ratio & $\tilde{Z} = \frac{\rho_{\rm{s}} c_{\rm{L}}}{\rho_{\rm{f}} c_0}$ & 1 & 8.4 & 0.7 & 1 & - \\
														\midrule
Wavelength  & $\lambda$ & \SI{}{\micro\metre}	& - & - & - & 600\\
Displacement amplitude 	& $u_{0}$ & \SI{}{\nano\metre} 	& - & - & - & 0.1 \\
On-resonance frequency & $f_\mathrm{res} = \frac{2w}{c_0}$ & MHz   & - & - & - & 1.24 \\
Off-resonance frequency & $f_{\mathrm{off}}$ & MHz & - & - & - &6.65 \\
														\midrule
\end{tabular}
\end{table} %

\subsection{Acoustic eigenmodes} \label{sect:Resonance}
Due to the high impedance ratio $\tilde{Z} = 8.4$ for pyrex relative to water, see Table~\ref{tab:Parameters}, it is possible in the full pyrex model to excite a resonance in the device at the frequency $f_\mathrm{res} = 2w/c_0 = 1.24$~MHz, which is close to the ideal standing half-wave pressure eigenmode of the reduced hard-wall system. At this resonance frequency, the pressure amplitude $\big|p_\mathrm{f}\big|$ in the water is several times larger than the pressure amplitude $\rho_\mathrm{f} c_0 \: \omega u_0$ set by the imposed SAW displacement, and the resonance field mainly depends on the frequency and not significantly on the detailed actuation along the boundary \cite{Muller2012}. The full pyrex model and the reduced hard-wall model are therefore expected to be in good agreement at $f_\mathrm{res}$, as is verified by the right column in \figref{AllHard}.

In contrast, at off-resonance frequencies, such as $f_\mathrm{off} = 6.65$~MHz in the left column of \figref{AllHard}, the detailed actuation does matter. The lower left panel of \figref{AllHard} is an example of this, as it highlights an aspect that restricts the validity of the reduced hard-wall model. For the full model with 60-\SI{}{\micro\metre}-thick pyrex walls, the maximum displacement along the top boundary of the water domain is approximately four times larger than the displacement amplitude $u_0$ of the imposed SAW boundary condition on the bottom boundary of the water domain. This indicates that the system is actuated close to a structural acoustic eigenmode of the pyrex. An amplification is also seen in the lower right panel of \figref{AllHard} although to a smaller degree. This amplification of boundary displacements brought about by the existence of structural eigenmodes is not taken into account in the reduced hard-wall model.

\subsection{Physical limitations of the lossy-wall condition} \label{sect:LossyDiscussion}
The comparison between the reduced lossy-wall model and the full PDMS model in \figref{4Lossy} shows a clear mismatch. The most important  reasons for this are that the lossy-wall model neglects the actuation of both the solid and fluid domain, and that it neglects the transverse motion of the PDMS along the PDMS-water interface.

As for the hard-wall model, the lossy-wall model neglects the strong direct transfer of acoustic energy from the SAW to the PDMS wall, and the implications are the same: the lossy-wall model underestimates the deformation and motion of the PDMS-water boundaries due to this. Moreover, due to the low impedance ratio $\tilde{Z} = 0.7$, there are no strong resonances in the water domain like the one at $f_\mathrm{res}$ for which the detailed boundary conditions do not matter.

In contrast to the reduced hard-wall model, some aspects of the deformation and motion of the PDMS-water boundaries are taken into account in the reduced lossy-wall model, as it includes the partial reflection and absorption waves from the water domain with perpendicular incidence on the PDMS wall. While this approach would be a good description of a planar or weakly curving interface between two fluids, where all the acoustic excitation takes place in one of the fluids, it is of limited use in the present system, for three reasons: (1) As discussed above, the acoustic energy is injected by the SAW into both the water and the PDMS domain. (2) The PDMS-water boundary is not planar, but consists of three linear segments joined at right angles. (3) PDMS is not a fluid, but supports shear waves, which are neglected in the reduced lossy-wall model. These three aspects are all part of the full PDMS model, in which PDMS is described as a linear elastic material supporting both longitudinal and transverse waves.

\subsection{Modeling PDMS as a linear elastic} \label{sect:ElasticsReason}
When modeling large strains above 0.4 in PDMS, non-linear effects are commonly included using hyperelasticity models in the form of a constitutive relation for the stress and strain for which the elastic moduli depends on the stress instead of being constant. For small strains below 0.4, PDMS becomes a usual linear elastic material  \cite{Kim2008, Schneider2008, Hohne2009, Still2013, Johnston2014}. In our model, the calculated displacements within the PDMS walls are less than 10~nm, corresponding to strains below $10^{-4}$, which justifies the use of linear elastics as the governing equations of the PDMS walls in our system. The use of linear elasticity is further validated in the literature, where linear elastic models of PDMS yield results comparable to those found when using more complex approaches, such as a Mooney--Rivlin constitutive model \cite{Yu2009}, a neo--Hookian approach \cite{Bourbaba2013}, and a Maxwell--Wiechert model \cite{Lin2009}.

Further simplifications based on neglecting the transverse motion of PDMS, such as modeling it as a fluid \cite{Leibacher2014, Darinskii2016} and applying the lossy wall conditions \cite{Nama2015}, are not advised, since PDMS does have a non-zero transverse bulk modulus and does support transverse sound waves \cite{Madsen1983, Still2013, Johnston2014}.

As characterization results for PDMS are scarce in the literature, we had to combine the material parameters found in Refs.~\cite{Madsen1983} and \cite{Zell2007} in our simulations.

%%%%%%%%%%%%%%%%%%%%%%%%%%%%%%%%%%%%%%%%%%

\section{Materials and Methods} \label{sect:Numeric}

Our modeling is based on the generic device design \cite{Shi2011,Guldiken2012} illustrated in \figref{Chip}. The device consists of a long, straight, fluid-filled microchannel surrounded by an elastic solid wall on the sides and top. The microchannel and walls rest on a piezoelectric substrate, along which a standing SAW is imposed as a boundary condition. We assume translational invariance along the axial $x$ direction, and only model the transverse $yz$ plane. We implement 2D numerical models in COMSOL Multiphysics~5.2 \cite{COMSOL52} using the parameters listed in Table~\ref{tab:Parameters}.  All acoustic fields are treated using an Eulerian description, and they have a harmonic time-dependence of the form $\vec{u}_\mathrm{s}(y,z)\:\ee^{-\iot}$, such that $\partt$ becomes $- \ii \omega$, where $\mathrm{i} = \sqrt{-1}$ while $\omega = 2 \pi f$ is the angular frequency and $f$ the frequency of the imposed SAW. For simplicity, we often suppress the spatial and temporal variable and write a field simply as $\vec{u}_\mathrm{s}$.

\begin{figure}[t]
\centering
\includegraphics[width=\columnwidth]{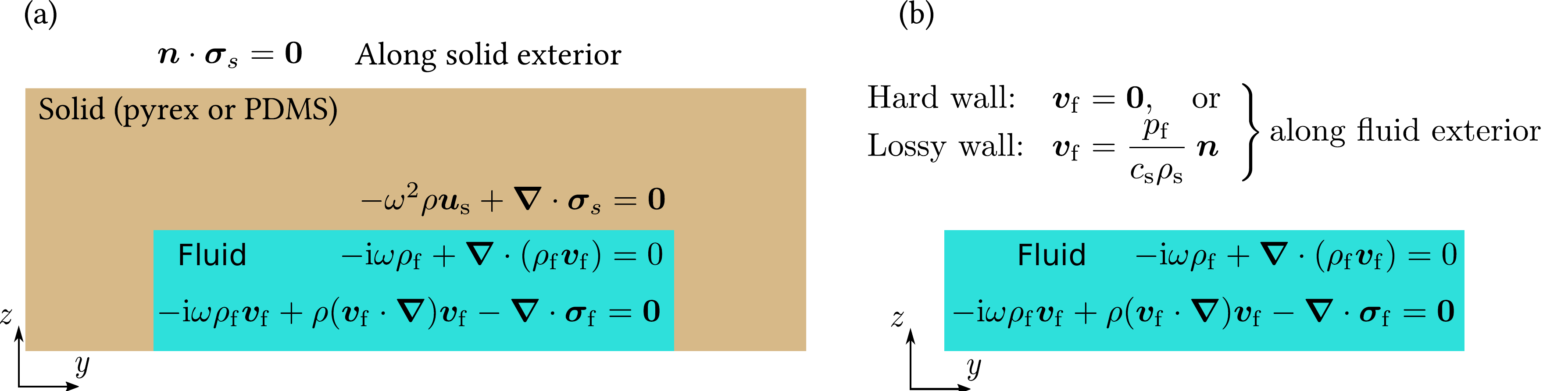}
\caption{\label{fig:models} Sketches of the models used in the study. (a) The full model with a solid domain (pyrex or PDMS) and a fluid domain (water). (b) The reduced model with only a fluid domain with boundary conditions (hard or lossy) representing the surrounding solid (pyrex or PDMS, respectively).}
\end{figure}

In total four models are set up, all with the imposed SAW as a boundary condition representing the actual piezoelectric lithium niobate substrate. (1) The full pyrex model, \figref{models}(a),  where the solid wall is modeled as a linearly elastic material with the parameters of pyrex, while the fluid is modeled as water. (2) The reduced hard-wall model, \figref{models}(b), where only the water is modeled, while hard-wall boundary conditions replace the pyrex wall. (3) The full PDMS model, \figref{models}(a), which is the full pyrex model in which the pyrex parameters are replaced by PDMS parameters. (4) The reduced hard-wall model, \figref{models}(b), where only the water is modeled, while lossy-wall boundary conditions replace the PDMS wall.

\subsection{Governing equations}
The unperturbed fluid at constant temperature $T = 298$~K in the fluid domain is characterized by its density $\rho_0$, viscosity $\eta_0$, and speed of sound $c_0$. The governing equations for the acoustic pressure $p_\mathrm{f}$, density $\rho_\mathrm{f}$, and velocity $\vec{v}_\mathrm{f}$ are the usual mass and momentum equations. The constitutive equation between the acoustic pressure $p_\mathrm{f}$ and density $\rho_\mathrm{f}$ is the usual linear expression, $p_\mathrm{f} = c_0^2\:\rho_\mathrm{f}$. Neglecting external body forces on the fluid, while applying perturbation theory \cite{Bruus2012} and inserting the harmonic time-dependence, the governing equations and the constitutive equation are linearized to following first-order expressions,

\begin{subequations} \label{eq:AcousticFirst}
\begin{align}
\label{eq:pertcont} \ii \omega p_\mathrm{f}  &=  \rho_0 c_0^2 \vec{{\nabla}} \cdot  \vec{v}_\mathrm{f},\\
\label{eq:pertCauchyF} -\rho_0 \ii \omega \vec{v}_\mathrm{f} &= \vec{\nabla} \cdot \ten{\sigma}_\mathrm{f},\\
\label{eq:stressF} \ten{\sigma}_\mathrm{f} &=
\eta_0 \Big[\vec{\nabla}\vec{v}_\mathrm{f} + \big(\vec{\nabla}\vec{v}_\mathrm{f} )^\mathrm{T}\Big]
+ \beta\eta_0\vec{\nabla}\big(\vec{\nabla}\cdot\vec{v}_\mathrm{f}\big)\:\vec{I},
\end{align}
\end{subequations}
where we have introduced the Cauchy stress tensor $\ten{\sigma}_\mathrm{f}$, and where superscript 'T' denotes tensor transpose, $\beta$ is the bulk-to-shear viscosity ratio, and $\vec{I}$ is the unit tensor. With appropriate boundary conditions, the first-order acoustic fields $p_\mathrm{f}$, $\rho_\mathrm{f}$, and $\vec{v}_\mathrm{f}$, can be fully determined by \eqref{AcousticFirst}. The specific model-dependent boundary conditions are presented and discussed in Sections~\ref{sect:Intro} and \ref{sect:Boundaryconditions}.

The dynamics in the solid of unperturbed density $\rho_\mathrm{s}$ is described by linear elastics through the momentum equation in terms of the displacement field $\us$ and the solid stress tensor $\sigs$. The constitutive equation relating displacement and stress is defined using the longitudinal $c_{\rm{L,s}}$ and transverse $c_{\rm{T,s}}$ speeds of sound of the given solid,

\begin{subequations} \label{eq:ElasticBasis}
\begin{alignat}{4}
\label{eq:CauchyS2} -\rho_\mathrm{s} \omega^{2} \us  &= \divv \sigs, \\
\label{eq:stressc}  \sigs &= \rho_\mathrm{s} \left[c_{\rm{T,s}} ^{2}(\grad \us  + \grad \us^{T}) + (c_{\rm{L,s}}^{2}-2c_{\rm{T,s}}^{2}) (\divv \us) \vec{I} \right].
\end{alignat}
\end{subequations}

\subsection{Boundary conditions} \label{sect:Boundaryconditions}
For simplicity the full dynamics of the piezoelectric substrate is not modeled. Instead, the standing SAW is implemented by prescribing displacements $\vec{u}_\mathrm{pz} = \big(u_{y,\mathrm{pz}}, u_{z,\mathrm{pz}}\big)$ in the $y$ and $z$ directions, respectively, on the bottom boundary of our domain using the following analytical expression form the literature \cite{Koster2007,Nama2015}, where the damping coefficient of $\SI{118}{\per\metre}$ has been neglected given the small dimensions ($< 0.002$~m) of the microfluidic device \cite{Nama2015},

 \begin{subequations} \label{eq:SAWdisp}
 \begin{align}
 u_{y,\mathrm{pz}} &= 0.6 u_0  \bigg\{\sin\!\Big[ k \big(\mbox{$\frac12$} w-y\big)+\omega t\Big]
 + \sin\!\Big[k \big(y-\mbox{$\frac12$} w\big)+\omega t \Big] \bigg\},\\
 u_{z,\mathrm{pz}} &=  -u_0 \bigg\{\cos\!\Big[k\big(\mbox{$\frac12$} w - y\big)+\omega t \Big]
 +\cos\!\Big[k (y- \mbox{$\frac12$} w)+\omega t \Big] \bigg\},\\
 \label{eq:BCfSAW} \vec{v}_\mathrm{f} &= -\ii \omega \vec{u}_\mathrm{pz},
   \quad  \text{imposed on the fluid at the fluid-SAW interface},\\
 \label{eq:BCsSAW} \vec{u}_\mathrm{s} &= \vec{u}_\mathrm{pz},
   \quad  \text{imposed on the solid at the solid-SAW interface}.
 \end{align}
 \end{subequations}
where $k =2\pi/\lambda$ is the wavenumber and $u_0$ the displacement amplitude of the SAW.

In the full models, a no-stress condition for  $\sig$ is applied along the exterior boundary of the solid. On the interior fluid-solid boundaries continuity of the stress is implemented as a boundary condition on $\ten{\sigma}_\mathrm{s}$ in the solid domain imposed by the fluid stress $\ten{\sigma}_\mathrm{f}$, while continuity of the velocity is implemented as a boundary condition on $\vec{v}_\mathrm{f}$ in the fluid domain imposed by the solid velocity $-\mathrm{i}\omega\:\vec{u}_\mathrm{s}$. Along the free surfaces of the solid a no-stress condition is applied,

 \begin{subequations} \label{eq:Fluid-solid}
 \begin{alignat}{3}
 \label{eq:BCfs} \vec{n}_{\mathrm{s}}\cdot\sigs &= \vec{n}_\mathrm{s}\cdot \sigf,
   \quad &&\text{imposed on the solid at the fluid-solid interface},\\
 \label{eq:BCsf} \vec{v}_\mathrm{f} &= - \ii \omega \us,
   \quad  &&\text{imposed on the fluid at the fluid-solid interface}, \\
 \label{eq:BCsFREE} \vec{n}_{\mathrm{s}}\cdot\sigs &= \vec{0}
   \quad  &&\text{imposed on the solid at exterior boundaries}.
 \end{alignat}
 \end{subequations}

In the reduced models, boundary conditions are imposed on the fluid to represent the surrounding material. Stiff and heavy materials such as pyrex are represented by the hard-wall (no motion) condition at the boundary of the fluid domain. Soft and less heavy materials such as PDMS are represented by the lossy-wall condition for partial acoustic transmittance perpendicular to the boundary of the fluid domain. For both conditions, a no-slip condition is applied on the tangential velocity component. The specific expression implemented in COMSOL are,

 \begin{subequations} \label{eq:WallConditions}
 \begin{alignat}{2}
 \label{eq:HardWall}  \vec{v}_\mathrm{f} &= \vec{0},
 \quad &&\text{boundary condition representing hard walls}, \\
 \label{eq:LossyWall} \vec{v}_\mathrm{f} &= \frac{p_\mathrm{f}}{c_s \rho_s} \: \vec{n},
 \quad &&\text{boundary condition representing lossy walls}.
 \end{alignat}
 \end{subequations}

\subsection{Numerical implementation and validation}
\label{sect:comsol}

\begin{figure}[t]
\centering
\includegraphics[width=\columnwidth]{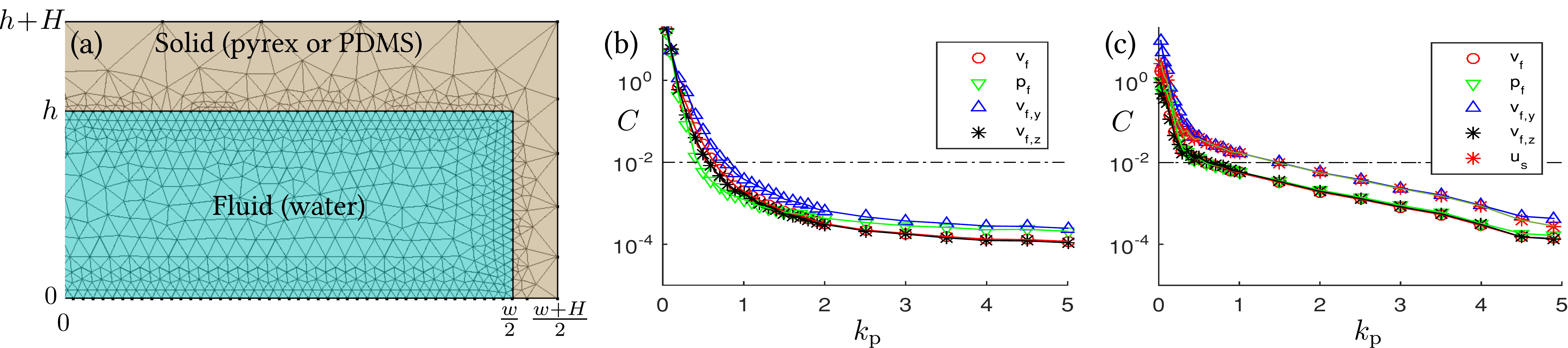}
\caption{\label{fig:mesh} (a) The mesh implemented in COMSOL, here shown in a coarse version for illustrative purposes with the small value $k_p = 0.1$ for the mesh parameter.  (b) For each of the fields $p_\mathrm{f}$, $\vec{v}_\mathrm{f}$, $v_{\mathrm{f},y}$, and $v_{\mathrm{f},z}$, the relative mesh convergence parameter $C$ is plotted versus mesh parameter $k_p$ for the reduced lossy-wall model. The dashed line represents $C = 0.01$.  (c) The same as in (b) but for the water domain in the full PDMS model with the inclusion of the field $\vec{u}_\mathrm{s}$.}
\end{figure}

We follow our previous work \cite{Muller2012, Muller2013}, and implement the governing equations in weak form in the commercial software COMSOL Multiphysics 5.2 \cite{COMSOL52}. To fully resolve the thin acoustic boundary layer of width $\delta$,

 \begin{equation}
 \delta = \sqrt{\frac{2\eta_0}{\rho_0\omega}} = \SI{0.21}{\micro\meter}\;
 \text{ at }\; \omega = 2\pi\times 6.5~\text{MHz},
 \end{equation}
in the water domain near its edges, the maximum mesh size $h_\mathrm{edge}$ at the solid-fluid boundary is much smaller than that in the bulk called $h_\mathrm{bulk}$. Both of these are controlled by the mesh parameter $k_\mathrm{p}$,

 \begin{equation}
 h_\mathrm{edge} = \frac{1}{k_\mathrm{p}}\:\delta,
 \quad
 h_\mathrm{bulk} = 50 h_\mathrm{edge},
 \quad
 \text{with }\;
 k_\mathrm{p} = 1\; \text{ in the main runs }.
 \end{equation}
The coarse mesh with $k_\mathrm{p} =0.1$ is shown in Fig.~\ref{fig:mesh}(a). In our largest (full) models using $k_\mathrm{p} = 1$, the implementation resulted in $8.1 \times 10^{6}$ degrees of freedom and a computational time of 30~minutes on a standard pc work station. The implementation of the model in the fluid domain has been validated both numerically and experimentally in our previous work \cite{Muller2012, Muller2013}. The solid domain implementation was validated by calculating resonance modes for a long rectangular cantilever, clamped at one end and free at the other, and comparing them successfully against analytically known results. Finally, for both the full and the reduced models, we performed a mesh convergence analysis using the relative mesh convergence parameter $C(g)$ for a given field $g(y,z)$ as introduced in Ref.~\cite{Muller2012},

  \begin{equation}
  C(g) = \sqrt{\frac{\int_\Omega(g-g_\mathrm{ref})^2\:\dm y \dm z}{\int_\Omega(g_\mathrm{ref})^2\:\dm y \dm z}}.
  \end{equation}
Here, $g_\mathrm{ref}$ is the solution obtained with the finest possible mesh resolution, in our case the one with mesh parameter  $k_\mathrm{p} = 5$. For all fields, our mesh analysis revealed that satisfactory convergence was obtained with the mesh parameter set to  $k_\mathrm{p} = 1$. For this value, the relative mesh convergence parameter was both small, $C \approx 0.01$, and exhibited an exponential asymptotic behavior, $C \simeq \ee^{-k_\mathrm{p}}$, as a function of the mesh parameter $k_\mathrm{p}$, for two examples see Fig.~\ref{fig:mesh}(b) and (c).

%%%%%%%%%%%%%%%%%%%%%%%%%%%%%%%%%%%%%%%%%%
\section{Conclusions}
A numerical method has been presented for 2D full modeling of a generic SAW microdevice consisting of a long, straight, fluid-filled microchannel encased in a elastic wall and resting on a piezoelectric substrate in which a low-MHz-frequency standing SAW is imposed. We have also presented reduced models consisting only of the fluid domain, where boundary conditions are used as simplified representations of the elastic wall. An acoustically hard wall, such as pyrex, is represented by a hard-wall boundary condition, while an acoustically soft wall, such as PDMS, is represented by a lossy-wall boundary condition. Our results show that the full pyrex model is approximated fairly well for thick pyrex walls using the hard-wall model, when the SAW is actuated on a frequency corresponding to a resonance frequency of the water domain, but less well for thinner walls at resonance and for any wall thickness off resonance. The reduced lossy-wall model was found to approximate the full PDMS model poorly, especially regarding the resulting running pressure waves in the reduced lossy-wall model in contrast the standing waves in the full PDMS model.

Modeling of acoustofluidic devices should thus be performed in full to take into account all effects relating to the elastic walls defining the microchannel. At higher frequencies or higher acoustic power levels, even the full model presented here must be extended to take into account thermoviscous effects in the form of increased heating and temperature-depending effects \cite{Muller2014, Ha2015}.

%%%%%%%%%%%%%%%%%%%%%%%%%%%%%%%%%%%%%%%%%%
\vspace{6pt}

%%%%%%%%%%%%%%%%%%%%%%%%%%%%%%%%%%%%%%%%%%
%% optional
%\supplementary{The following are available online at www.mdpi.com/link, Figure S1: title, Table S1: title, Video S1: title.}

%%%%%%%%%%%%%%%%%%%%%%%%%%%%%%%%%%%%%%%%%%
%\acknowledgments{All sources of funding of the study should be disclosed. Please clearly indicate grants that you have received in support %of your research work. Clearly state if you received funds for covering the costs to publish in open access.}

%%%%%%%%%%%%%%%%%%%%%%%%%%%%%%%%%%%%%%%%%%
\authorcontributions{N.R.S and H.B. contributed equally to the work}

%%%%%%%%%%%%%%%%%%%%%%%%%%%%%%%%%%%%%%%%%%
\conflictofinterests{The authors declare no conflict of interest.}
%Declare conflicts of interest or state ``The authors declare no conflict of interest.'' Authors must identify and declare any personal circumstances or interest that may be perceived as inappropriately influencing the representation or interpretation of reported research results. Any role of the funding sponsors in the design of the study; in the collection, analyses or interpretation of data; in the writing of the manuscript, or in the decision to publish the results must be declared in this section. If there is no role, please state ``The founding sponsors had no role in the design of the study; in the collection, analyses, or interpretation of data; in the writing of the manuscript, and in the decision to publish the results''.}

%%%%%%%%%%%%%%%%%%%%%%%%%%%%%%%%%%%%%%%%%%
%% optional
\abbreviations{The following abbreviations are used in this manuscript:\\

\noindent
%MDPI: Multidisciplinary Digital Publishing Institute\\
%DOAJ: Directory of open access journals\\
%TLA: Three letter acronym\\
%LD: linear dichroism \\
Pyrex: Borosilicate glass \\
PDMS: Polydimethylsiloxane \\
SAW: Surface acoustic wave \\
BAW: Bulk acoustic wave \\
IDT: Interdigital transducer
}

%%%%%%%%%%%%%%%%%%%%%%%%%%%%%%%%%%%%%%%%%%
%% optional
%\appendix
%\section{}
%The appendix is an optional section that can contain details and data supplemental to the main text. For example, explanations of experimental details that would disrupt the flow of the main text, but nonetheless remain crucial to understanding and reproducing the research shown; figures of replicates for experiments of which representative data is shown in the main text can be added here if brief, or as Supplementary data. Mathemtaical proofs of results not central to the paper can be added as an appendix.
%
%\section{}
%All appendix sections must be cited in the main text. In the appendixes, Figures, Tables, etc. should be labeled starting with `A', e.g., Figure A1, Figure A2, etc.

%%%%%%%%%%%%%%%%%%%%%%%%%%%%%%%%%%%%%%%%%%
\bibliographystyle{mdpi}

%=====================================
% References, variant A: internal bibliography
%=====================================
\renewcommand\bibname{References}

%\begin{thebibliography}{999}
%% Reference 1
%\bibitem{ref-journal}
%Lastname, F.; Author, T. The title of the cited article. {\em Journal Abbreviation} {\bf 2008}, {\em 10}, 142-149.
%% Reference 2
%\bibitem{ref-book}
%Lastname, F.F.; Author, T. The title of the cited contribution. In {\em The Book Title}; Editor, F., Meditor, A., Eds.; Publishing House: City, Country, 2007; pp. 32-58.
%\end{thebibliography}

%=====================================
% References, variant B: external bibliography
%=====================================
%\bibliography{acoustofluidics}

%%%%%%%%%%%%%%%%%%%%%%%%%%%%%%%%%%%%%%%%%%
%% optional
%\sampleavailability{Samples of the compounds ...... are available from the authors.}

%%%%%%%%%%%%%%%%%%%%%%%%%%%%%%%%%%%%%%%%%%
\end{document}